\documentclass[prb,twocolumn,superscriptaddress,floatfix]{revtex4}
\usepackage{hyperref}
\usepackage{graphicx} 
\usepackage{amsmath}
\usepackage[dvips]{epsfig}
\newcommand{\beq}{\begin{equation}}
\newcommand{\eeq}{\end{equation}} 
\newcommand{\beqa}{\begin{eqnarray}}
\newcommand{\eeqa}{\end{eqnarray}}
\newcommand{\ba}{\begin{array}}
\newcommand{\ea}{\end{array}}

\begin{document}

\title{
Coexistence of vortex arrays and surface capillary waves in spinning prolate superfluid $^4$He nanodroplets}

\author{Mart\'{\i} Pi}
\affiliation{Departament FQA, Facultat de F\'{\i}sica,
Universitat de Barcelona, Av.\ Diagonal 645,
08028 Barcelona, Spain.}
\affiliation{Institute of Nanoscience and Nanotechnology (IN2UB),
Universitat de Barcelona, Barcelona, Spain.}

\author{Jos\'e Mar\'{\i}a Escart\'{\i}n}
\affiliation{Catalan Institute of Nanoscience and Nanotechnology (ICN2),
CSIC and BIST, Campus UAB,
Bellaterra,
08193 Barcelona, Spain.}

\author{Francesco Ancilotto}
\affiliation{Dipartimento di Fisica e Astronomia ``Galileo Galilei''
and CNISM, Universit\`a di Padova, via Marzolo 8, 35122 Padova, Italy.}
\affiliation{CNR-IOM Democritos, via Bonomea, 265 - 34136 Trieste, Italy.}

\author{Manuel Barranco}
\affiliation{Departament FQA, Facultat de F\'{\i}sica,
Universitat de Barcelona, Av.\ Diagonal 645,
08028 Barcelona, Spain.}
\affiliation{Institute of Nanoscience and Nanotechnology (IN2UB),
Universitat de Barcelona, Barcelona, Spain.}

\begin{abstract} 
Within Density Functional Theory, we have studied  the interplay between vortex arrays and capillary waves in spinning prolate 
$^4$He droplets made of several thousands of helium atoms. Surface capillary waves are ubiquitous in prolate  superfluid $^4$He droplets and,
depending on the size and angular momentum of the droplet, they may coexist with vortex arrays. We have found that 
the equilibrium configuration of small prolate droplets is vortex-free, evolving towards vortex-hosting as the droplet size increases.		
This result is in agreement with a recent experiment [S.~M.~O. O'Connell {\it et at.}, Phys.\ Rev.\ Lett.\ \textbf{124}, 215301 (2020)], where it has 
been disclosed that  vortex arrays and capillary waves coexist in the equilibrium configuration of very large drops.
 Contrarily to viscous droplets executing rigid body rotation, the stability phase diagram  of spinning $^4$He droplets cannot be universally
described in terms of dimensionless angular momentum  and angular velocity variables:
instead,
the rotational properties of superfluid helium droplets display a clear dependence on the droplet size and the number of vortices they host.
 \end{abstract} 
\date{\today}

\maketitle

\section{Introduction}
\label{sec1}

Superfluid $^4$He droplets produced in the expansion of a cold helium gas,\cite{Toe04} or by hydrodynamic instability
of a liquid helium jet passing through the nozzle of a molecular beam apparatus,\cite{Tan18} 
have been considered  the ultimate  inert matrix for molecular  spectroscopy,\cite{Leh98,Cal11}  constituting also 
an ideal testground to study superfluidity at the nanoscale.\cite{Sin89,Kri90,Gre98,Kro01,Tan02,Bra13,Mat15}   
A series of experiments aiming at determining the appearance of large spinning $^4$He droplets made of 
$10^8-10^{11}$ atoms\cite{Gom14,Jon16,Ber17,Rup17,Lan18,Ges19,Oco20} 
has motivated a renewed  interest in aspects such as
how vortices are distributed inside helium droplets,\cite{Anc15,Anc18} how impurities are captured 
by the vortex lines they host, and  how impurities arrange inside vortex-hosting droplets.\cite{Gom12,Lat14,Vol15,Cop17,Bla18,Cop19,Ern21}  

Two experimental techniques have been used to address large spinning $^4$He droplets. The pioneering work of Ref.\ \onlinecite{Gom14}
and other works carried out by the same group\cite{Jon16,Ber17,Oco20}
used coherent diffractive imaging of x-rays from a free electron laser (FEL) and gave access to a model-independent determination 
of the two-dimensional (2D) projection of the droplet density on a plane perpendicular to the x-ray incident direction via iterative phase retrieval algorithms;
by doping He droplets with Xe atoms
they were also able to detect the presence of vortex arrays.
In Refs.\ \onlinecite{Rup17,Lan18}, He droplets were irradiated with intense extreme ultraviolet pulses (XUV), 
and the measurements of wide-angle diffraction patterns
provided access to full three-dimensional information. However, it was not possible
to directly determine the shapes of the droplets, and so the analysis was carried out parameterizing the
droplet density with a combination of two ellipsoidal caps 
connected by a hyperboloidal centerpiece, and all their defining parameters were
determined by matching the experimental diffraction patterns with those obtained from simulations.

These remarkable experiments have shown that most helium droplets are spherical and only a few percent are deformed. 
In particular, the analysis  carried out by Langbehn \textit{et al.}\cite{Lan18} 
on a sample consisting  of a large number of droplets (38{,}150) showed that 92.9\% of them were spherical, 5.6\% were
spheroidal (oblate\cite{note}), and 1.5\% were prolate. The latter population was made of ellipsoidal  (0.8\%),   pill-shaped (0.6\%), and 
dumbbell-like (0.1\%) droplets, where the appearance was 
quantitatively determined by evaluating the distance of the center of mass 
of the droplet to its surface along the direction of the principal axes of inertia.   
The results obtained with both experimental techniques  were compared to calculations made for incompressible viscous droplets only subject 
to surface tension  and centrifugal  forces;\cite{Bro80,Hei06,But11,Bal15} it was concluded that they were in agreement. 

The precise meaning of this agreement needs some clarification, as otherwise one might be left with the impression that 
the rotational properties of superfluid $^4$He and of classical rotating viscous fluids are 
very similar, 
and this is not the case.\cite{Lam45,Guy15} The experiments have indeed  shown that the 
shapes of spinning superfluid $^4$He drops are the same as those found for viscous drops subject to centrifugal and 
surface tension forces, 
and that some relationships between ratios of the mentioned distances are in agreement with 
the classical results,\cite{Bal15} see, {\it e.g.}, Fig.\ 3 of Ref.\ \onlinecite{Lan18} and  Fig.\ \ref{fig1} below.  However, to ascertain whether the
sequence of shapes is the same, which is what determines the equilibrium phase diagram, it is necessary 
to experimentally determine the angular 
momentum for a large enough sample of droplets; this has not been possible yet. 
Only very recently, the angular momentum and shape of a 
few He drops have been simultaneously determined.\cite{Oco20} 

At first glance, the mere fact that {\it oblate} $^4$He spinning drops have 
been identified\cite{Gom14,Lan18,Oco20} is surprising and unexpected. Large helium drops
are produced in the normal phase and may acquire angular momentum 
during the passage of the fluid through the nozzle of the 
molecular beam apparatus. The drops cool down to a temperature of 0.37~K\cite{Toe04} 
and are thus superfluid when they are probed.
The natural question is how a spinning superfluid drop may acquire an oblate (axisymmetric) shape as 
this is quantum mechanically forbidden.
In fact the equations characterizing the macroscopic behaviour of a superfluid 
at zero temperature have the form of irrotational hydrodynamics, from which the 
moment of inertia can be calculated as\cite{Cop17,Boh75,Pit16}
$\Theta_{\rm irr} = \varepsilon^2 \, \Theta _{\rm rig}$, where  
\begin{equation}
\varepsilon =\frac{\langle y^2-x^2 \rangle}{\langle y^2+x^2 \rangle}
\label{eq61}
\end{equation}
and 
$\Theta _{\rm rig}=N m \langle x^2+y^2\rangle$ is the rigid moment of inertia, $N$ being the
number of atoms in the droplet and $m$ the atomic mass, 
showing that in a superfluid the value of the moment of inertia 
is smaller than the rigid value. 
In particular, for axisymmetric systems (i.e. $ \langle x^2 \rangle =\langle y^2 \rangle$) the above relation
gives for the angular momentum of the superfluid
$\langle L_z \rangle =\Theta_{\rm irr} \, \omega =0$.
Notice that these results 
are not restricted to the perturbative regime of small $\omega $.\cite{Boh75}
These expressions show that vortex-free oblate axisymmetric configurations 
cannot spin, whereas prolate (non axisymmetric)
configurations can, and the resulting angular momentum  
$L_{\rm cap} =  \Theta_{\rm irr} \, \omega$ is associated with 
the so-called capillary waves (see the discussion in the following).

Thus, the only likely explanation for the experimental observation 
of spinning oblate $^4$He droplets
is that these drops contain a vortex array which
can store most of the angular momentum
acquired during the passage through the experimental apparatus
--some is taken away by the atoms emitted during the cooling process. 
It has indeed been shown\cite{Anc15,Anc18}  that 
the presence of a large enough vortex array in the  
$^4$He droplets confers to them a globally oblate appearance,\cite{note}
similarly to the case of rotating viscous droplets.
Vortex arrays have been experimentally detected only in a few  
cases,\cite{Gom14,Jon16,Ber17,Oco20}  as they escape detection 
unless drops are doped and the presence of vortices is 
established by the appearance of Bragg patterns from the impurities that fill 
the vortex cores. 

The existence of helium drops that follow the  sequence of prolate shapes 
characteristic of viscous drops under the effect of rotation is  also worth
discussing. In this case, axial symmetry is spontaneously broken 
($\langle x^2 \rangle  \ne  \langle y^2 \rangle$), and the droplet can store a finite amount
of angular momentum in the form of capillary waves, even in the absence of vortices.
In general, however, capillary waves and vortices may coexist in 
spinning prolate droplets, producing 
two quite different irrotational velocity flows 
compatible with the superfluidity requirement: one associated to quantized vortices,
and the other to surface capillary waves.\cite{Boh75,Don91,Pit16,Sei94}  
Remarkably, in the stability phase diagram, spinning prolate 
superfluid drops whose angular momentum is exclusively stored in capillary waves lie on 
a line\cite{Anc18} quite distinct from the classical line,\cite{Bro80} as discussed in the following. 

There is another unexpected difference between a rotational viscous 
fluid and an irrotational superfluid with finite angular momentum.
When a spherical viscous droplet is set into rotation, 
as a consequence of centrifugal forces it flattens out in the direction of  
the rotational axis: the shortest 
half-axis of the droplet ($c_z$, as defined in the following)
is thus aligned with the rotational axis.
However, this is not always the case for irrotational flows:
spinning vortex-free droplets with prolate shapes
have instead the intermediate half-axis ($b_y$, as defined in the following)
aligned with the rotational axis.\cite{Lam45}
When vortices are present,  in most cases
the shortest half-axis $c_z$
is again aligned with the rotational axis,
in agreement with the classical behavior.
Consequently, if only the shape of the droplet is experimentally identified,
the rotational axis cannot be determined unambiguously if no further information
about the angular momentum vector is available.

From the above discussion it appears that capillary waves and vortex arrays can 
coexist in prolate superfluid drops.
This has been demonstrated in a recent combined experimental and Density Functional Theory (DFT) work.\cite{Oco20} 
The DFT analysis was restricted to deformable self-sustained superfluid $^4$He cylinders  which  
could host a fairly large vortex array due to the lower dimensionality (2D) of the computational real space grid.
This model was a generalization of  the rotating, elliptic cylindric vessel filled with $^4$He studied in Ref.\ \onlinecite{Fet74}.     
 
In this work we complement the discussion made in Ref.\ \onlinecite{Oco20} by addressing several prolate configurations corresponding
to large  (for DFT standards)  superfluid $^4$He droplets:  we put the emphasis on the interplay between vortices and
capillary waves, we determine the equilibrium configuration for two representative 
values of the angular momentum, and we discuss  the similarities and 
differences between superfluid (irrotational) and viscous (rotational) behavior. 
Oblate droplets hosting a different number of vortices have been thoroughly studied in Ref.~\onlinecite{Anc15}.

This paper is organized as follows: the DFT approach to liquid helium is recalled in Sec.\ \ref{sec2}; Sec.\ \ref{sec3} presents
the results and their discussion; and a summary is given in Sec.\ \ref{sec4}. Multimedia information is presented in the Supplemental Material.\cite{SM}

\section{Theoretical Approach}
\label{sec2}
 
Density Functional Theory in its static and time-dependent  (TDDFT) versions has proved to be  a powerful method
to address superfluid helium droplets.\cite{Bar06,Anc17}
Within DFT, the energy of the $^4$He droplet is written as a functional of the atom density $\rho({\mathbf r})$ as
\begin{equation}
E[\rho] = T[\rho] + E_c[\rho] =
\frac{\hbar^2}{2m} \int d {\mathbf r} |\nabla \Psi({\mathbf r})|^2 +  \int d{\mathbf r} \,{\cal E}_c[\rho]
\label{eq1}
\end{equation}
where the first term  is the kinetic energy, with $\rho({\mathbf r})= |\Psi({\mathbf r})|^2$, and 
the functional ${\cal E}_c$ contains the interaction term (in the
Hartree approximation) and additional terms which describe non-local correlation effects.\cite{Anc05}
  
The droplet equilibrium configuration is obtained by solving the Euler-Lagrange (EL) equation resulting 
from the functional variation of Eq.\ (\ref{eq1}),
\begin{equation}
\left\{-\frac{\hbar^2}{2m} \nabla^2 + \frac{\delta {\cal E}_c}{\delta \rho}  \right\}\Psi({\mathbf r}) 
 \equiv {\cal H}[\rho] \,\Psi({\mathbf r})  = \mu \Psi({\mathbf r})
\;,
\label{eq3}
\end{equation}
where $\mu$ is the $^4$He chemical potential corresponding to the number of He atoms in the droplet, 
$N = \int d{\bf r}|\Psi({\bf r})|^2$. 

To study spinning $^4$He droplets  it is convenient 
to work in the fixed-droplet frame of reference (corotating frame at  angular velocity $\omega$), {\it i.e.}, we consider
\begin{equation}
E'[\rho] = E[\rho] - \hbar \omega \, \langle \hat{L}_z \rangle  
\label{eq5}
\end{equation}
where $\hat{L}_z$ is the dimensionless angular momentum operator in the $z$-direction; 
one looks for solutions of the EL equation resulting from the functional variation of  $E'[\rho]$,
\begin{equation}
\left\{{\cal H}[\rho] \,-\hbar \omega \hat{L}_z\right\} \,\Psi(\mathbf{r})  =  \,\mu \, \Psi(\mathbf{r})
\;.
\label{eq7}
\end{equation}
The above equation has been solved using the $^4$He-DFT-BCN-TLS 
computing package,\cite{Pi17} see  Refs.~\onlinecite{Anc17} and \onlinecite{dft-guide} and references therein for additional details.
Briefly, we work in cartesian coordinates, with the effective wave 
function $\Psi(\mathbf{r})$ defined at the nodes of a 3D grid inside a
calculation box large enough to accommodate the droplet in such a way 
that the He density is sensibly zero on the box surface. Periodic
boundary conditions are imposed which allow to use the Fast Fourier 
Transform to carry out the convolutions needed to obtain the DFT
mean field ${\cal H}[\rho]$. The differential operators 
in ${\cal H}[\rho]$ and $\hat{L}_z$ are approximated by 13-point
formulas and the typical space step is
0.8 \AA.
Working at fixed angular momentum requires to iterate on the value of $\omega$; 
there are  efficient ways of adjusting iteratively $\omega$, {\it e.g.},
the Augmented Lagrangian method.\cite{Gar21}

\begingroup
\squeezetable
\begin{table}
\begin{center}
\begin{tabular}{|c|c|c|c|c|c|c|}
\hline
\multicolumn{7}  {|c|} {$\Lambda=0.80$  \quad $L= 4155$ \quad $L_{\rm cap}= 4014$ \quad  $N=1500$ } \\
\hline
 $n_v$ & $E$  (K) & $E_{\rm kin}$ (K) & $\omega$ (K/$\hbar$)  & $a_x$ {\rm (\AA)}& $b_y {\rm (\AA)}$ & $c_z {\rm (\AA)}$\\
\hline
0    & -7909.33    & 419.82 & 0.1145  & 42.38 & 15.93 &18.93\\
1    & -7916.16     & 478.10 & 0.1269  & 39.75 & 18.53 & 18.07 \\
{\bf 2}    & {\bf -7918.89}     & {\bf 533.96} & {\bf 0.1349}  & {\bf 36.50} & {\bf 20.71} & {\bf 19.32} \\
3(T)    & -7905.11    & 594.92& 0.1431   & 32.76 & 24.13 &19.56 \\
4(C)    &  -7885.24   & 635.61&  0.1315 & 28.69 & 28.39 & 20.32 \\
\hline
\end{tabular}
\end{center}
\caption{
Energetics and morphology of  prolate $^4$He$_{1500}$  droplets   at $\Lambda=0.80$ for $n_v=0-4$.
$T$ indicates a triangular and $C$ a cross-shaped vortex array. The entry in boldface corresponds to the 
globally stable configuration. 
\label{Table1}
}
\end{table}
\endgroup

Angular momentum can be stored in a superfluid $^4$He droplet 
in the form of surface capillary waves and/or quantized vortices.\cite{Sei94,Anc18,Oco20} 
In order to deposit angular momentum of both types in the droplet, 
we have used the so-called ``imprinting'' procedure by starting the imaginary time minimization from a very flexible guess
 for the  effective wave function 
$\Psi(\mathbf{r})$, namely a superposition of a quadrupolar capillary wave and $n_v$ vortex lines parallel to the $z$ axis,
\begin{equation}
\Psi_0(\mathbf{r})=\rho_0^{1/2}(\mathbf{r})\,  e^{\imath\,\alpha x y} \,
\prod _{j=1}^{n_v} {(x-x_j)+\imath (y-y_j) \over \sqrt{(x-x_j)^2+(y-y_j)^2}}
\;,
\label{eq57}
\end{equation}
which is then optimized by iteratively solving Eq.~(\ref{eq7}). Here, $\rho_0(\mathbf{r})$ is an arbitrary, vortex-free droplet 
density, the complex phase $e^{\imath \alpha x y}$  imprints a surface capillary wave with quadrupolar symmetry 
around the $z$ axis,\cite{Cop17} and the product term imprints a vortex array made of $n_v$ linear vortices,\cite{Anc17}
where $(x_j, y_j)$ is the initial position of the $j$-vortex core.
The initial value of $\alpha$ and the vortex core positions are guessed, and
during the iterative solution of Eq.~(\ref{eq7})
both the vortex core structure and positions, and the droplet shape change
to provide at convergence the lowest energy configuration.

\begingroup
\squeezetable
\begin{table}
\begin{center}
\begin{tabular}{|c|c|c|c|c|c|c|}
\hline
	\multicolumn{7}  {|c|} {$\Lambda=0.80$  \quad $L=85317$ \quad $L_{\rm cap}= 82449$ \quad  $N=20{,}000$ } \\
\hline
 $n_v$ & $E$  (K) & $E_{\rm kin}$ (K) & $\omega$ (K/$\hbar$)  & $a_x$ {\rm (\AA)}& $b_y {\rm (\AA)}$ & $c_z {\rm (\AA)}$\\
\hline
0    & -127236.24    &  2319.50 & 0.03172  & 100.07 &  37.75 & 44.90 \\
1    & -127320.36    &  2495.08 & 0.03422  & 96.22  &  41.65 & 44.75  \\
2    & -127397.02    &  2654.06 & 0.03607  & 91.70  &  44.51 & 45.97 \\
{\bf 3(L)}    & {\bf -127432.28}     & {\bf 2831.78} & {\bf 0.03733}  & {\bf 87.11} & {\bf 47.88} & {\bf 44.73} \\
4(C)   & -127428.73    &  3021.48 & 0.03908  & 82.20  & 52.50  & 46.62 \\
4$^*$(L)   & -127405.10    &  3027.91 & 0.03745  & 82.85  & 51.37  & 46.43 \\
5(P)    & -127413.79    &  3214.34 & 0.04021  & 75.66  & 58.05  & 47.34 \\
\hline
\end{tabular}
\end{center}
\caption{
Energetics and morphology of  prolate $^4$He droplets  with $N=20{,}000$  at $\Lambda=0.80$ for $n_v=0-5$.
$L$ indicates a linear,  $C$ a cross-shaped, and $P$ a pentagonal vortex array. The entry in boldface corresponds to the 
globally stable configuration, and the asterisk  identifies the highest energy   $n_v=4$ configuration.
\label{Table2}
}
\end{table}
\endgroup

Writing $\Psi(\mathbf{r}) \equiv \phi(\mathbf{r})\, \exp[i\,\cal{S}(\mathbf{r})]$, the velocity field of the superfluid is
\begin{equation}
\mathbf{v}(\mathbf{r}) = \frac{\hbar}{m} {\rm Im}\left\{\frac{\nabla \Psi(\mathbf{r})}{\Psi(\mathbf{r})}\right\}
= \frac{\hbar}{m} \nabla \cal{S}(\mathbf{r})
\label{eq577}
\end{equation}
The velocity potential $\cal{S}(\mathbf{r})$ gets contributions from both
capillary waves and vortices in a complex way, and their contributions
cannot be rigorously disentangled ---the analysis needs to be model-dependent.\cite{Oco20} 

\begingroup
\squeezetable 
\begin{table}
\begin{center}
\begin{tabular}{|c|c|c|c|c|c|c|}
\hline
	\multicolumn{7}  {|c|} {$\Lambda=0.80$ \quad  $L= 163899$ \quad $L_{\rm cap}=158398$ \quad  $N=35{,}000$ } \\
\hline
 $n_v$ & $E$  (K) & $E_{\rm kin}$ (K) & $\omega$ (K/$\hbar$)  & $a_x$ {\rm (\AA)}& $b_y$ {\rm (\AA)}& $c_z$ {\rm (\AA)} \\
\hline
0       & -227393.50  & 3360.11 & 0.02402  & 120.49 & 45.42 & 54.21 \\
1       & -227522.50  & 3581.02 & 0.02580  & 116.29 & 49.68 & 54.09 \\
2       & -227629.67  & 3767.38 & 0.02697  & 111.55 & 52.80 & 55.29 \\
3       & -227698.13  & 3986.17 & 0.02789  & 106.56 & 56.28 & 53.89 \\
4(C)    & -227721.40  & 4233.29 & 0.02927  & 101.38 & 61.12 & 56.03 \\
4$^*$(L)    & -227700.04  & 4233.98 & 0.02829  & 101.76 & 60.15 & 55.82 \\
{\bf 5(P)} & {\bf -227723.95} & {\bf 4476.43}& {\bf 0.03009}  & {\bf 95.16} & {\bf 66.16 } & {\bf 56.86} \\
6(Ct)  & -227699.01 & 4732.32 & 0.03130 & 85.12 & 75.48 & 53.36 \\
6$^*$       & -227697.29 & 4710.33 & 0.03012 & 87.06 & 73.60 & 57.59 \\
\hline
\end{tabular}
\end{center}
\caption{
Energetics and morphology of  prolate $^4$He droplets with $N=35{,}000$ at $\Lambda=0.80$ for $n_v=0-6$.
$L$ indicates a linear,  $C$ a cross-shaped, and $P$ a pentagonal vortex array.  
The 6($Ct$) configuration has a center vortex, {\it i.e.},  along the rotational axis.
The entry in boldface corresponds to the  globally stable configuration, and
the asterisk  identifies the highest energy configuration among those with the same $n_v$ value; in particular,
6$^*$ corresponds to a  6-vortex array configuration in which the six vortices are at the vertices of a hexagon.
\label{Table3}
}
\end{table}
\endgroup

The existence of different prolate configurations for a $^4$He$_N$ droplet spinning at a given $L$, 
each of them characterized by a number  of vortices  $n_v$, raises the question  of
which is the globally stable configuration and how does it depend on the number of atoms in the droplet. 
To determine which is the globally stable configuration among different ones, 
one has to compare their total energy $E[\rho]$ including the rotational energy  (Routhian).\cite{Bro80,Hei06}
At variance with the classical case, where a rotational energy term  
in the rigid body approximation has to be added to the energy of 
the droplet,\cite{Bro80,But11} in the DFT approach to  superfluid $^4$He 
in the corotating frame this is naturally accounted for through 
the velocity field embodied in the phase of the effective wave  function $\Psi({\mathbf r})$, see Eq. (\ref{eq577}),  
so one does not need to add any extra term to the total energy expression, Eq. (\ref{eq1}) --nor to Eq. (\ref{eq7}). 

\begingroup
\squeezetable 
\begin{table}
\begin{center}
\begin{tabular}{|c|c|c|c|c|c|c|}
\hline
\multicolumn{7}  {|c|} {$\Lambda=1.25$ \quad  $L= 6493$ \quad $L_{\rm cap}= 6224$ \quad  $N=1500$ } \\
\hline
 $n_v$ & $E$  (K) & $E_{\rm kin}$ (K) & $\omega$ (K/$\hbar$)  & $a_x$ {\rm (\AA)}& $b_y$ {\rm (\AA)}& $c_z$ {\rm (\AA)} \\
\hline
{\bf 0}    & {\bf -7658.10}     & {\bf 519.11}& {\bf 0.1001}  & {\bf 48.41} & {\bf 13.63} & {\bf 15.40} \\
1    & -7641.16    &  583.36 & 0.1025   & 46.68 & 15.80 & 14.52 \\
2    & -7626.09    & 648.30 &  0.1153   & 44.44 & 17.15 & 16.33 \\
3(L)    & -7601.71     &  714.49& 0.1197  &42.51 & 18.72 & 15.93 \\
4(L)    &  -7567.76  &  763.02 &  0.1222 &  42.13 & 19.28 & 16.86 \\
\hline
\end{tabular}
\end{center}
\caption{
Energetics and morphology of  prolate $^4$He$_{1500}$ droplets  at $\Lambda=1.25$ for $n_v=0-4$.
$L$ indicates a linear vortex array. The entry in boldface corresponds to the  globally stable configuration. 
\label{Table4}
}
\end{table}
\endgroup
\begingroup
\squeezetable
\begin{table}
\begin{center}
\begin{tabular}{|c|c|c|c|c|c|c|}
\hline
\multicolumn{7}  {|c|} {$\Lambda=1.25$ \quad $L = 133307$ \quad $L_{\rm cap}= 127937$ \quad  $N=20{,}000$ } \\
\hline
 $n_v$ & $E$  (K) & $E_{\rm kin}$ (K) & $\omega$ (K/$\hbar$)  & $a_x$ {\rm (\AA)}& $b_y$ {\rm (\AA)}& $c_z$ {\rm (\AA)}\\
\hline
{\bf 0}    & {\bf -125806.09}   & {\bf 2883.61} & {\bf 0.02781}  & {\bf 114.26} & {\bf 32.39} & {\bf 36.54}  \\
1    & -125789.56  & 3097.10 & 0.02953  & 111.73 & 35.85 & 36.16 \\
2    & -125789.99  & 3293.39 &   0.03089  & 108.35 & 37.43  &  38.51 \\
3    & -125767.52  & 3494.95 &  0.03203 &  105.28 & 39.80 & 38.28  \\
4$^*$(C)& -125701.35 & 3676.09 &0.03306 &  103.20 & 42.47 & 38.92 \\
4(L) & -125714.08  & 3705.57 &  0.03283 &  102.53 & 41.98 & 39.20 \\
5    & -125641.07  & 3887.18 &  0.03371 &  100.35 & 44.39 & 39.88 \\
\hline
\end{tabular}
\end{center}
\caption{Energetics and morphology of  prolate $^4$He droplets with $N=20{,}000$  at $\Lambda=1.25$ for $n_v=0-5$.
$L$ indicates a linear, and $C$ a cross-shaped vortex array. The entry in boldface corresponds to the  globally stable configuration,  
and the asterisk  identifies the highest energy $n_v=4$ configuration.
\label{Table5}
}
\end{table}
\endgroup

A difficulty inherent  to the study of helium droplets is that configurations with rather distinct morphologies may have similar energies;
therefore,
a careful analysis is required in order to distinguish the global minimum from metastable configurations.
An added challenge is that most vortex array configurations are very robust and,
in spite of not being a physically conserved  quantity, the  starting $n_v$ value 
is often conserved instead of being relaxed to the optimal
value corresponding to the global  equilibrium configuration.
This forces to explore, for a given $N$-atoms droplet, all possible $n_v$ values compatible with
the chosen $L$ ---a task which becomes increasingly cumbersome as $L$ increases.  
This robustness has however the  benefit that one has access to a series of excited configurations,
which might be experimentally accessible,
characterized not only by the strictly conserved $N$ and $L$ values but also by the number of vortices  $n_v$. 
We have always checked the stability of the vortex array  against shape distortions.

\begingroup
\squeezetable
\begin{table}
\begin{center}
\begin{tabular}{|c|c|c|c|c|c|c|}
\hline
\multicolumn{7}  {|c|} {$\Lambda=1.25$ \quad $L = 256093$ \quad $L_{\rm cap}= 245756$ \quad  $N=35{,}000$ } \\
\hline
 $n_v$ & $E$  (K) & $E_{\rm kin}$ (K) & $\omega$ (K/$\hbar$)  & $a_x$ {\rm (\AA)}& $b_y$ {\rm (\AA)}& $c_z$ {\rm (\AA)}\\
\hline
0 & -225315.35 &  4174.71& 0.02102   & 137.65 & 38.98 & 43.97 \\
1 & -225307.23&4450.00 & 0.02225  & 134.88 & 42.80 & 43.50\\
{\bf 2} &{\bf  -225323.77} & {\bf 4695.91} & {\bf 0.02318} & {\bf 131.14} & {\bf 44.40} & {\bf 46.30} \\
3 & -225312.04 &4950.83 & 0.02402 & 127.81 & 47.07 & 46.14 \\
4$^*$(C) & -225239.14 & 5198.51& 0.02477 &  125.33 & 50.06 & 46.70 \\
4(L) & -225263.35& 5217.47& 0.02464 &  124.69 & 49.46 & 47.12 \\
5(P) & -225185.14& 5466.69& 0.02529  &122.15 &52.15 & 47.86 \\
5$^*$(L) & -225173.77 & 5489.12 &0.02505  & 122.16 & 51.69 & 45.90 \\
6 & -225104.71 & 5713.58 & 0.02600 & 118.84&  54.80 &  48.43\\
\hline
\end{tabular}
\end{center}
\caption{Energetics and morphology of  prolate $^4$He droplets with $N=35{,}000$  at $\Lambda=1.25$ for $n_v=0-6$. 
$L$ indicates a linear,  $C$ a cross-shaped, and $P$ a pentagonal vortex array. The  6-vortex array configuration is a stretched hexagon.
The entry in boldface corresponds to the  globally stable configuration, and 
the asterisk  identifies the highest energy configuration among those with the same $n_v$ value. 
\label{Table6}
}
\end{table}
\endgroup

\begin{figure}[!]
\centerline{\includegraphics[width=1.0\linewidth,clip]{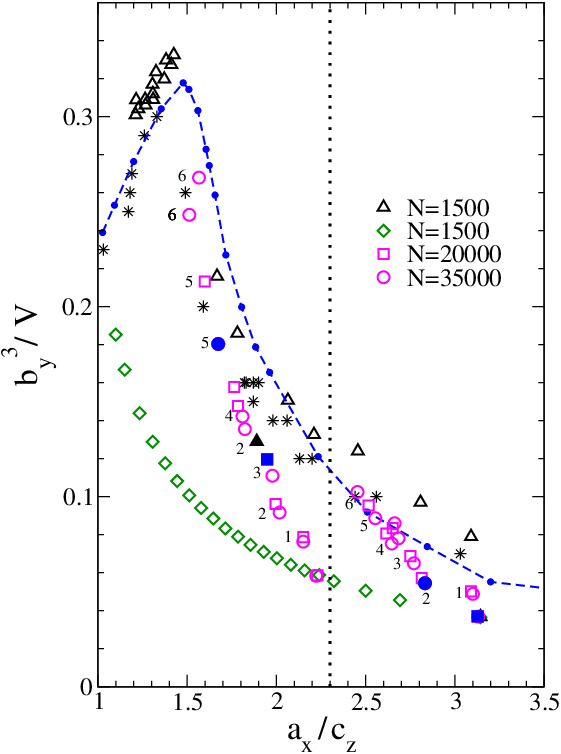}}
\caption{
Aspect-ratio  $b_y^3/V$ {\it vs.}   $a_x/c_z$ curve.
Black triangles: 3-vortex $^4$He$_{1500}$ configurations.\cite{Anc18}  Green diamonds: vortex-free (capillary wave)
$^4$He$_{1500}$ configurations.\cite{Anc18}
Magenta  squares (circles): $N=20{,}000 \,(35{,}000)$ configurations. The numbers close to the symbols indicate the $n_v$ value.
The vertical dotted line  separates the $N=20{,}000$ and 35{,}000 droplet configurations with $\Lambda=0.8$ (left) from those with  $\Lambda=1.25$ (right).
Solid symbols represent the global equilibrium configurations displayed in the tables.
Black starred symbols are the experimental results of Ref. \onlinecite{Lan18}, and
blue dots connected by a dashed line are the classical rotating drop results.\cite{But11,Bal15}
The inset displays the lengths $a_x, b_y$  and $c_z$ used to characterize the droplet shape.
}
\label{fig1}
\end{figure}

For every stationary configuration obtained by solving Eq.~(\ref{eq7}), a   ``sharp density surface'' is determined by calculating
the locus at which the helium density equals $\rho_0/2$, where  $\rho_0$ is the atom density of the  liquid; for a
spherical distribution this corresponds to the sphere of radius $R$ defined below. 
In the case of deformed droplets, three lengths (half-axes) $a_x$, $b_y$ and $c_z$ 
are introduced  representing the distances from the  center of mass of the droplet to the sharp surface along the principal axes of inertia. 
These lengths are represented in the inset in Fig. \ref{fig1}.
For an axisymmetric droplet, $a_x=b_y \ne c_z$. These lengths have been used  to characterize the droplet shape by defining two 
dimensionless aspect ratios,   $a_x/c_z$ and $b_y^3/V$,\cite{Bal15,Anc18,Lan18} where $V$ is the volume of the non-rotating
spherical droplet.

\begin{figure}[!]
\centerline{\includegraphics[width=1.0\linewidth,clip]{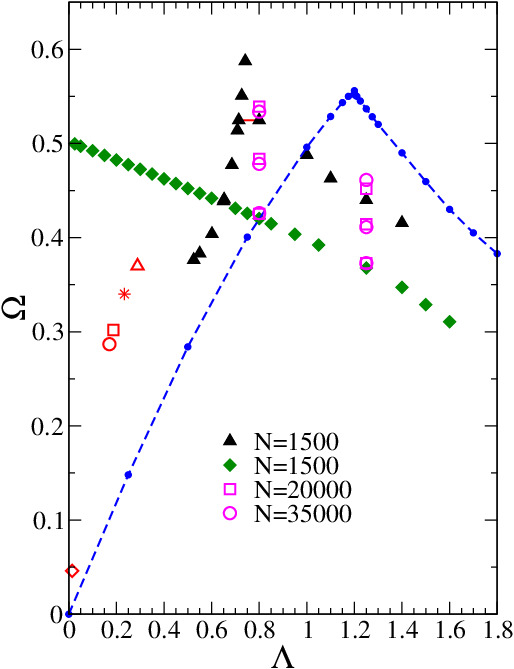}}
\caption{
Dimensionless angular velocity $\Omega$ \textit{vs}.\ dimensionless angular momentum $\Lambda$.
Black triangles: 3-vortex $^4$He$_{1500}$ configurations.\cite{Anc18}  Green diamonds: vortex-free (capillary wave) $^4$He$_{1500}$
configurations.\cite{Anc18}
The oblate and prolate configurations connected by a red horizontal line at $\Omega= 0.5247$ are the
twin configurations we refer to in the text. 
Magenta squares (circles) represent the $N=20{,}000 \, (35{,}000)$ results for $n_v=0, 2$, and 5 configurations (for  given $N$ and $\Lambda$ values, 
the larger the $n_v$, the bigger the $\Omega$).
For the sake of completeness, we also include the critical points for nucleating a single vortex (red symbols):
triangle, $N=1500$; asterisk, $N=5000$;\cite{Cop19} square, $N=20{,}000$; circle, $N=35{,}000$.  
The red diamond at $(\Lambda,\Omega)=(1.4\times 10^{-2}, 4.6 \times 10^{-2})$ is critical point for nucleating a vortex for $N= 10^9$
in the hollow core model.
The blue dots connected by a dashed line are the classical rotating drop results.\cite{But11,Bal15}
}
\label{fig2}
\end{figure}
\begin{figure*}[t] 
\centerline{\includegraphics[width=1.0\linewidth,clip]{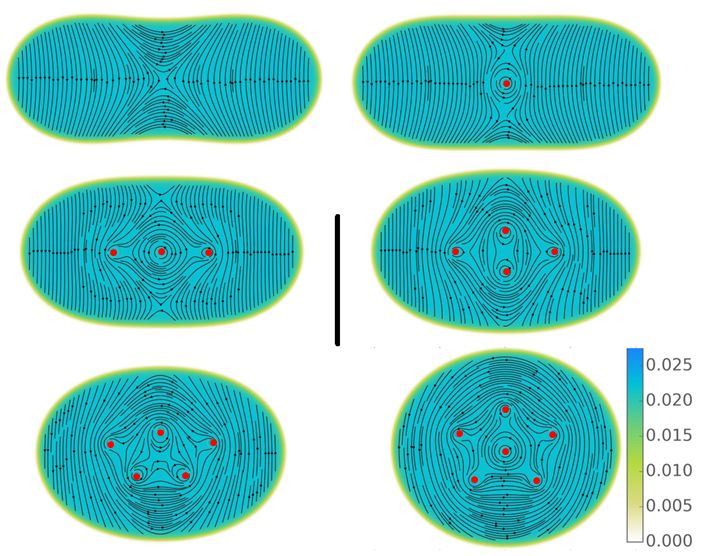}}
\caption{
Densities on the symmetry plane perpendicular to the rotational axis
for the $N=35{,}000$ He droplet hosting zero, one, and  and $n_v= 3-6$ vortices at $\Lambda=0.80$.
The streamlines of the superfluid flow are superimposed and their direction is  counterclockwise. 
The positions of the vortex cores are marked by red dots for visualization. The color bar shows the atom density in units of 
\AA$^{-3}$, and the black  bar represents a distance of 100 \AA. 
}
\label{fig3}
\end{figure*}

The shapes of classical drops {\it exclusively} subject to surface tension and centrifugal forces 
follow a universal stability diagram in terms of 
the dimensionless  angular momentum $\Lambda$ and angular velocity $\Omega$ defined by\cite{Bro80,But11} 
\begin{equation}
\begin{split}
 \Omega &\equiv \sqrt{\frac{m \, \rho_0 \, R^3}{8 \, \gamma}}\;  \omega 
\\
\Lambda &\equiv \frac{\hbar}{\sqrt{8 \gamma R^7  m\rho_0}} \, L
\end{split}
\label{eq9}
\end{equation}
where $m$ is the atom mass  and $\gamma$ is the surface
tension of the liquid. For liquid $^4$He at zero temperature and pressure,
 $\gamma$ = 0.274 K \AA$^{-2}$, $\rho_0$ = 0.0218 \AA$^{-3}$, the  sharp radius 
of the spherical  droplet  is given by $R= r_0 N^{1/3}$ with  $r_0= 2.22$ \AA,\cite{Bar06},
and  $\hbar^2/ m$ = 12.12 K \AA$^2$. We recall that $\hbar= 7.6382$ K\,ps and that $\hbar c= 2.2899 \times 10^7$ K \AA.

We want to stress that this universality is lost if one goes  beyond that simple model, {\it e.g.}, by using a DFT description
of  the liquid, since this incorporates effects such as surface diffuseness and liquid incompressibility:\cite{Pi20} although such effects 
are expected to be less important as the droplet size 
increases, they can be tangible for the 
droplet sizes investigated here. 
Yet, Eqs.~(\ref{eq9}) are very useful as they allow for a comparison between experimental and calculated 
droplets, even if the latter are much smaller 
(for obvious computational reasons) than those probed in the experiments. 
However, one should have  in mind that some differences may appear in the 
comparison due to these unavoidable finite-size effects. 
Besides, the presence of vortices in the case of superfluid droplets definitely  breaks the universality
of the stability diagram, as shown below.

\section{Results and discussion} 
\label{sec3}

Systematic DFT calculations are very cumbersome to carry out, especially when several configurations 
hosting different numbers of vortices have to be analyzed and their geometries have to be tested against distortions of the vortex array. This hinders
a systematic exploration of the phase diagram. For this reason, we have limited our study to some relevant cases which embody the physics we aim to
discuss. We have focused in particular on two $\Lambda$ 
values that embrace a fairly large range of  prolate  configurations,\cite{Anc18} namely $\Lambda=0.80$ and 1.25, 
for   droplets with $N=1500$ atoms (a value for which a rather detailed series of  DFT calculations is available\cite{Anc18}) and with   
$N=20{,}000$ and 35{,}000; these  large prolate droplets (for DFT standards) are well
suited to disclose the interplay between capillary waves and vortex arrays object of this study, and how the number of vortices
in the globally stable configuration  evolves as a function of $N$.  

Tables \ref{Table1}--\ref{Table6} collect the results obtained for the  mentioned $(\Lambda,N)$ values (one pair of values per table) and give details on their 
energetics and morphology.  In each table we have written in boldface the entry corresponding to the globally stable
configuration, {\it i.e.}, that with the lowest energy for the given values of $N$ and $\Lambda$.
When more than one configuration has been found for a given number of vortices $n_v$, the more energetic one is  identified with 
an asterisk ($^*$). Triangular, cross-shaped, pentagonal and linear vortex arrays are denoted by $T, C, P,$ and  $L$, respectively.
Notice that the  configurations in a given table can be directly
compared, as they correspond to the same values of two strictly conserved quantities in isolated droplets. 

As already discussed in the Introduction Section, it appears from the Tables
that for {\it vortex-free} superfluid droplets
the intermediate half-axis $b_y$ is aligned with the rotational axis,
at variance with the classical rotating droplets where 
the shortest half-axis $c_z$ is aligned instead with the rotational axis.
We are not aware of a general demonstration of this counter-intuitive result; 
its proof is however known for ellipsoidal 
drops made of irrotational (potential) fluids.\cite{Lam45}
This peculiarity of spinning irrotational drops has to be taken into account for a proper analysis of the experiments:
identifying the shortest axis of a spinning superfluid $^4$He droplet with the rotational  axis might be incorrect.
For {\it vortex-hosting} superfluid droplets, though, the rotational axis coincides most of the times with the shortest ($c_z$) one, 
as for viscous drops. 
But there are exceptions, {\it e.g.}, for  $N=20{,}000$ and 35{,}000 when $n_v=2$. 

In most cases, the presence of a vortex array confers to the 
droplet the appearance of a rotating viscous droplet. A notorious 
example of this apparently classical behavior 
is the meniscus that develops, at the liquid-vapor interface, 
in a rotating  bucket filled with superfluid helium above the critical
angular velocity required for vortex nucleation.\cite{Osb50,Don91} Hence, determining the angular momentum 
of the droplet (magnitude and direction) and whether it hosts a vortex array or not seems unavoidable before drawing a definite conclusion about 
how  superfluid helium droplets rotate.  
\begin{figure*}[t] 
\centerline{\includegraphics[width=1.0\linewidth,clip]{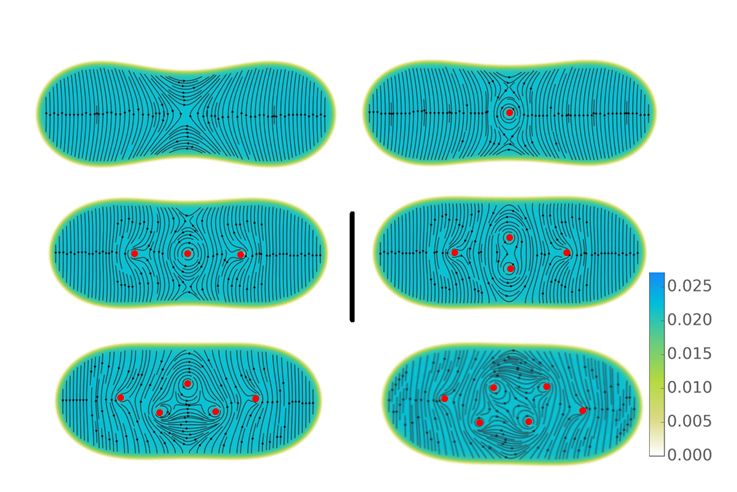}}
\caption{
Densities on the symmetry plane perpendicular to the rotational axis for the $N=35{,}000$ He droplet hosting  zero, one,  and
$n_v=3-6$  vortices at $\Lambda=1.25$. The streamlines of the superfluid flow are superimposed and their direction is 
counterclockwise.  The positions of the vortex cores are marked by red dots for visualization. The color bar shows the atom density 
in units of  \AA$^{-3}$, and the black  bar represents a distance of 100 \AA. 
}
\label{fig4}
\end{figure*}

We show in Fig.~\ref{fig1}  the dimensionless ratios $b_y^3/V$  versus  $a_x/c_z$ for the configurations collected in the tables,
irrespective of whether they correspond to globally stable or metastable configurations.
The $N=1500$ results have been complemented with those shown in  Fig.~1 of Ref.~\onlinecite{Anc18}, which presents data for a large 
sample of $\Lambda$ values. Also shown are 
 the experimental results,\cite{Lan18} which have an average atom number 
$\langle N \rangle = 6 \times 10^9$, and   the result
for classical rotating drops.\cite{Bal15}  
While the results for vortex-hosting configurations are similarly distributed and roughly follow the classical drops trend,
the figure shows that vortex-free prolate configurations form a separate branch (capillary branch)
displaying a completely different  behavior. Both branches meet at the $(a_x/c_z,b_y^3/V) = (1,3/4\pi)$ point corresponding to spherical droplets.

It is worth noticing that configurations with the same $n_v$ but rather different $N$ values yield similar points on the 
$(a_x/c_z, b_y^3/V)$ plane, especially when the vortex array has the same morphology (linear, cross, $\ldots$), and in particular 
for vortex-free configurations. 
The universality of the classical results, which allows to scale them with $N$, is not completely lost. Unfortunately, 
even $N=35{,}000$ is too small a value compared to  the experimental ones to allow us to draw any sensible conclusion
about this issue.
The figure also shows that, for a fixed $\Lambda$ value,  the larger the $n_v$
the more similar viscous and vortex-hosting droplets are.

We thus see that knowledge of the shape of a large number of deformed configurations is not enough to 
unambiguously characterize their rotational behavior as classical or superfluid,
and that one has to combine this information with the simultaneous determination of the angular momentum
of the droplet by, {\it e.g.}, obtaining the stability diagram in the $(\Lambda, \Omega)$ plane. This diagram is shown in Fig.~\ref{fig2} for 
some of the configurations in Fig.~\ref{fig1}. While the capillary branch may arrive up to nearly $\Lambda=0$, as only a small
deformation is needed for the droplet to sustain capillary waves, the vortex-hosting branch must abruptly interrupt at 
some $(\Lambda_c,\Omega_c)$ point, as
a critical angular velocity $\omega_c$ is needed for nucleating a vortex line in a superfluid droplet.\cite{Pit16}  The red symbols in
Fig. \ref{fig2} represent the critical vortex nucleation point for the three $N$ values used in this study. For the sake 
of completeness, we also display the DFT result for $N=5000$,\cite{Cop19} and that of
the hollow core model\cite{Dal00} for a $N= 10^9$ drop taking for the
vortex core radius a value of 1\AA. 

The existence of this critical point dramatically distinguishes  superfluid from viscous droplet rotational behavior, as
no axisymmetric  superfluid He droplet can be set into rotation for $\Lambda<\Lambda_c$.
Another conspicuous difference between viscous and superfluid vortex-hosting droplet results can be observed in Fig. \ref{fig2}, namely 
the location of the oblate-to-prolate bifurcation point. For viscous droplets, it is at 
$\Lambda=1.2$,\cite{But11,Hei06} whereas for droplets of the size studied here, DFT yields the bifurcation slightly below  $\Lambda = 0.8$. 
Indeed, the three-vortex $^4$He$_{1500}$ configurations connected by a red horizontal line correspond to  oblate (left)
and a prolate (right) configurations. 
We attribute this large difference to a finite size effect,
although we cannot ascertain this.  In Fig.~\ref{fig2}, finite size effects also clearly affect the position of the critical vortex nucleation point.

Comparing the results for the same $N$ value, one can see from the tables that  the ratio $b_y/a_x$ decreases 
as $\Lambda$ increases. This causes,
{\it e.g.}, that for the $^4$He$_{1500}$ droplet,  the  most stable $n_v=4$ vortex array, that was cross-shaped at $\Lambda=0.80$, is instead linear at
$\Lambda=1.25$; the same happens for $N=20{,}000$. 
Large droplets can host both kinds of vortex arrangements, see, {\it e.g.}, Table~\ref{Table6}.
For a given $\Lambda$, a large droplet may accommodate a non-linear
vortex array more easily.
Notice for instance that, at $\Lambda=1.25$, the cross-shaped $n_v=4$ configuration for $N=35{,}000$ is still more stable than the linear one.
The existence of multiple configurations for the same $n_v$ value is also ubiquitous in axisymmetric configurations,\cite{Cam79,Anc15,Anc14}  
making the search for the lowest energy configuration very time-consuming. 

Further inspection of the tables shows that, for given $N$ and $\Lambda$ values, the globally stable 
spinning prolate configurations are not necessarily  vortex-free, and that, contrarily to a naive expectation, 
for large, fixed $N$ and $\Lambda$ values, the energy is not  monotonously varying with $n_v$, see for instance 
Tables \ref{Table1}, \ref{Table2} and \ref{Table6}.  We have  found  that for the largest $\Lambda$ value 
considered in this study ($\Lambda=1.25$), the vortex-hosting configuration becomes the globally stable configuration only above a ``critical''
 $N$ value. We have checked that for  $N=30{,}000$ the vortex-free and the $n_v=2$ configurations are nearly degenerate with  energies of
 $-191\,994.86$~K and $-191\,994.49$~K, respectively. As  Table \ref{Table6} shows,  the globally stable configuration of the 
$N=35{,}000$ droplet is the $n_v=2$ one. It is thus very likely that the prolate large $N$ drops studied in the FEL\cite{Gom14} and 
XUV\cite{Lan18} experiments contain both vortex arrays and capillary waves.\cite{Oco20} Notice also that the energy differences between
the first excited configurations and the globally stable ones are small, and it cannot be completely discarded that some of such excited configurations are also detected
in the experiments.
These considerations indicate that vortex-free prolate configurations might be extremely difficult to detect either because they correspond  
to globally stable configurations only for small-$N$ droplets that escape detection in current
diffractive imaging experiments, or because they 
are metastable configurations in the case of large-$N$ drops and 
likely decay to vortex-hosting configurations before they are probed. 

Figures \ref{fig3} and \ref{fig4} show some 2D densities corresponding to the $N=35{,}000$ droplet.     
One may see that, in order to store a large angular momentum, a vortex-free configuration has to be very stretched
--see the discussion on capillary waves below.  
This is a general trend that implies an increase of the droplet surface energy; as shown in the tables,
the kinetic energy also increases with the number of vortices. 
The globally stable configuration is determined by a delicate balance between these two competing effects.
Figure \ref{fig4} shows that the $N=35{,}000$ droplet is large enough to accommodate up to 6 vortices, and that the shape of the
droplet  evolves from two-lobed  to a more compact, ``pill-shaped'' appearance as
found in the experiments.\cite{Ber17,Lan18,Oco20}  Despite most of these configurations
are metastable, these shape transitions exemplify the 
competition between compact and more linear vortex arrays arrangements to determine the appearance of the droplet. 
Notice that Fig. \ref{fig1} also displays  the same effect: as $n_v$  increases for a given $\Lambda$, the point representing
the vortex-hosting configuration dramatically displaces to smaller $a_x/c_z$ values, yielding  at the same time a larger $b^3_y/V$,
hence more compact droplet shapes. 
It is also worth seeing from the 2D figures that two-lobed configurations as those reported in Refs.~\onlinecite{Ber17} and \onlinecite{Lan18}
appear when $n_v$ is small; increasing $n_v$  reduces the wrist between the lobes,  as the vortex array 
becomes less linear and more similar to a patch of a triangular arrangement (as expected when $n_v$ becomes 
very large). 
Since the number of pill-shaped droplets found in the experiment\cite{Lan18}  is six times 
larger than that of dumbbell-like droplets, this would again indicate that large, very prolate droplets host fairly large vortex arrays.  
 
\begin{figure}[!] 
\centerline{\includegraphics[width=1.0\linewidth,clip]{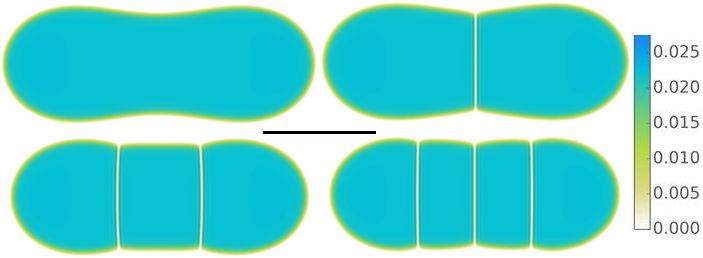}}
\caption{
Densities on the symmetry plane containing the vortex lines for the $N=35{,}000$ droplet hosting $n_v= 0-3$ vortices at $\Lambda=1.25$.
The color bar shows the atom density in units of  \AA$^{-3}$, and the black  bar represents a distance of 100 \AA.
}
\label{fig5}
\end{figure}

Figure \ref{fig5} displays a side view of the $N=35{,}000$ droplet at $\Lambda=1.25$ for the $n_v= 0-3$ configurations.
One can see how vortices locally distort the droplet surface as they have to hit it perpendicularly.  A
similar figure can be found in Ref. \onlinecite{Anc15} for oblate configurations. 
Thus, droplets hosting large vortex arrays are far from being ellipsoidal. In the diffractive imaging experiments, the images used to identify
the presence of vortices with a well oriented direction of their cores\cite{Oco20} would correspond to the flat surfaces limiting from above and below 
the droplet side view shown in this figure.

Vortex arrays in prolate configurations are distorted, as found in the experiments\cite{Oco20} and displayed in the 2D figures.
To characterize these distortions, we have calculated an oblate configuration ``twin'' of the
prolate $N=1500$, $\Lambda=0.80$, $n_v=3$ configuration for the same value of  $\Omega$, namely 0.5247,
which implies a lower $\Lambda$ value, see Fig. \ref{fig2}.  
For this $n_v=3$  oblate configuration we have found $\Lambda=0.714$, $a_x/c_z=1.38$, and $b_y^3/V=0.318$, 
the latter two values to be compared to 
those of its twin prolate configuration, $a_x/c_z=1.68$ and $b_y^3/V=0.207$.  
In the oblate $n_v=3$ configuration the vortex array  forms an equilateral triangle with an inter-vortex  distance of
$16.6$~\AA{}. In the $n_v=3$ prolate configuration the triangle is isosceles and 
stretched in the  direction of the largest principal axis of inertia ($x$ axis). The stretched inter-vortex distance is 22.1~\AA{}, and the other two 
sides of the triangle correspond to an inter-vortex distance of 15.4~\AA{}.

All configurations displayed in the 2D figures are stationary  in the corotating frame; 
as a consequence,  they would be seen from the laboratory frame
as if they  were rotating, apparently like a rigid-body, with the angular frequency $\omega$ imposed to obtain them.
 Let us recall and stress that this is a misleading appearance.
The motion of a fluid is a combination of translation, rotation and deformation of the
fluid elements, and only when the vorticity (defined as he curl of the velocity field of the fluid) is non zero, one may speak of a true rotation,
being important to distinguish between rotation and motion of the fluid elements along a curved path,\cite{Tri82,Don91}
{\it e.g.}, around a vortex core. Since the velocity field of the superfluid is potential, vorticity is zero 
 except on the vortex lines; in their absence, vorticity is zero everywhere.  
 
 \begin{figure}[!]
\centerline{\includegraphics[width=1.0\linewidth,clip]{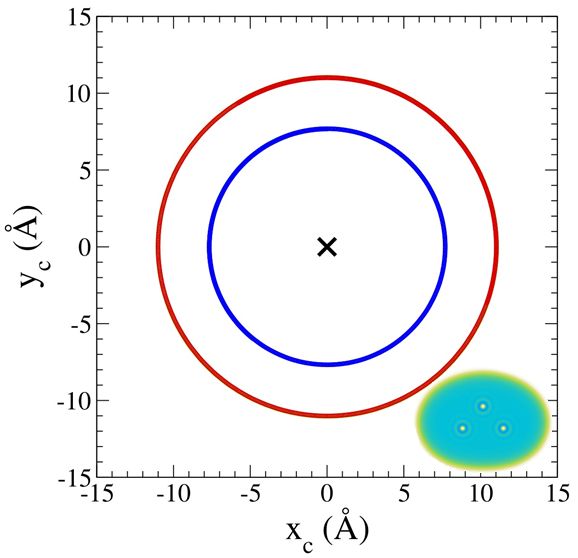}}
\caption{Trajectories of the vortex cores $(x_c,y_c)$ on the $z=0$ 
plane corresponding to the prolate $N=1500$, $\Lambda=0.80$, $n_v=3$ configuration.
One vortex is turning around the center of mass of the droplet (shown by a cross) 
at a distance of 7.7~\AA{}, while the other two vortices rotate at a distance of 11.0~\AA{}.
The inset represents the droplet density on the symmetry plane perpendicular to the rotational axis.
}
\label{fig6}
\end{figure}

 To illustrate this situation,  we have simulated within the TDDFT approach\cite{Anc17}
  the time evolution  of  the prolate $^4$He$_{1500}$ droplet for  the vortex-free and the $n_v=3$ configurations at 
 $\Lambda=0.8$.  Fig. \ref{fig6} shows, for the  $n_v=3$ case, the trajectory $(x_c,y_c)$ of the  
 vortex cores  on the $z=0$ plane.
It can be observed that the inter-vortex distances do not change during the evolution, 
indicating that the three-vortex array rotates with the angular  velocity imposed to the
corotating frame, $\omega = 1.871 \times 10^{-2}$ ps$^{-1}$. We see this dynamics as the rigid rotation of the vortex cores (which is not
in contradiction with the previous discussion, as they are empty), around which the irrotational superflow accommodates 
in its displacement. One may also notice the accuracy of the stationary solution we have obtained; otherwise, the
vortex core trajectories, plotted here point by point,  would not be the perfect circumferences displayed in  Fig. \ref{fig6}. 
The figure also exemplifies the robustness of the fixed-$n_v$ configurations mentioned in Sec. \ref{sec2}.
These configurations --which are stationary in the corotating frame--
likely correspond to  local energy minima and are  separated  by an energy barrier from 
other configurations having a different $n_v$ but the same $N$ and $\Lambda$ values.

The interested reader may find two videos  in the Supplemental Material\cite{SM}  showing the time evolution of these 
two configurations in the corotating frame for about 450~ps. It may be illuminating to look at the time evolution of the streamlines pattern: it
shows how different the {\it apparent} rotation of the superfluid droplet is, compared to the true rotation of a viscous drop.  
We want to stress that the energy, particle number and angular momentum of the isolated droplet are strictly conserved during the evolution. 

It is worth recalling that vorticity in quantum fluids may often lead to a turbulent regime\cite{Tsu09,Nem13} like, {\it e.g.},
that resulting from the decay of vortex tangles in bulk liquid $^4$He, or 
the turbulence in Bose Einstein condensates (BEC) generated by fast stirring the condensate using laser beams.
Our vortex structures, however, are stable against decay, reconnections or other mechanisms
which are known to lead to turbulent behavior in liquid $^4$He or BECs.
This is what we indeed observe in our simulations: in spite of the 
possibility of bending and displacing of the vortex cores, which is allowed by the three-dimensional
nature of our calculations, the vortex structures we find are either stationary states in the co-rotating system or, when let to evolve
under real-time dynamics, they rotate around the central axis but remain 
otherwise stable, without showing any tendency to be expelled or reconnect with nearby vortices. 
This is in agreement with the experimental measurements on 
nanoscopic $^4$He droplets hosting vortices where 
no sign of decay of the vortex structures is observed.\cite{Oco20}
Turbulent behavior may arise though in nanoscopic $^4$He droplets under 
special conditions, as illustrated in Ref. \onlinecite{Esc19}, where a TDDFT study
of the merging of two vortex-free superfluid $^4$He droplets has shown
that vortex rings are dynamically nucleated during the merging process
and that their annihilation produces a massive emission of rotons and 
subsequent turbulent behavior, whose scaling follows
the very general power laws underlying turbulence in 
quantum fluid systems.\cite{Tsu09,Nem13}   

Streamlines of the superfluid flow are shown in Figs. \ref{fig3} and \ref{fig4}. The superflow
follows the direction of the angular frequency imposed to  the corotating frame (counterclockwise in this case). 
Streamlines allow to infer by visual inspection
the coexistence of vortices and surface capillary waves, as their velocity fields are very different.
The streamlines associated to vortices wrap around their cores, whereas those associated  to  capillary waves 
end abruptly at the droplet surface.\cite{Fet74,Cop17,Oco20} 
As already mentioned, separating the contribution of capillary waves and vortices
in the velocity fields is generally not possible,
the superfluid velocity field being proportional to the gradient of the phase $\cal{S}(\mathbf{r})$ of 
the effective wave function $\Psi({\mathbf r})$,  where the contributions of both vortices and  capillary waves are entangled.
 
The coexistence of one single vortex and a quadrupolar irrotational flow in an elliptic cylinder filled with helium
rotating at constant $\omega$  was studied  by  Fetter,\cite{Fet74} who anticipated the possible appearance of a 
row of vortices along the longer axis of the elliptic vessel cross section. At variance with isolated droplets, the 
cylinder has a rigid surface of known geometry. Writing the phase  of
$\Psi({\mathbf r})$  as the sum of a term arising from the vortex line and another from the quadrupolar flow (the same guess that we 
made in Eq.~(\ref{eq57}), but in our case only for the starting configuration), he could  determine how the angular momentum is shared 
between the vortex  and the quadrupolar irrotational flow. It does not look feasible to extend the cylinder approach
to the $n_v>1$ case.
Even for one single vortex, the study of isolated droplets is much more complex, as their shapes are not ellipsoidal in general and they are 
only known after the stationary configuration has been determined by solving Eq.~(\ref{eq7}). 

Yet, to analyze the experimental results\cite{Oco20} it has been useful  to know, at least approximately,  how angular 
momentum is shared between vortices and capillary waves.  
In order to do that, we computed the deformation parameter $\varepsilon$ defined in Eq. (\ref{eq61})
using the $a_x$ and $b_y$ parameters given in the tables for the $n_v=0$ configurations,
and the relation giving the angular momentum
associated to capillary waves, 
$L_{\rm cap} =  \varepsilon^2 \, \Theta _{\rm rig} \, \omega$ (see the discussion following Eq. (\ref{eq61})).

We have found that it works surprisingly well, as it coincides with the exact $L$ calculated within the DFT approach to within a few percent.
The $L_{\rm cap}$ value obtained for each studied $n_v=0$ configuration is given in the tables. 
For vortex-hosting configurations, the angular momentum stored in the vortex arrays may then be estimated subtracting $L_{\rm cap}$ 
from the total $L$. The validity of this procedure cannot be assessed though. 
So far, the only possibility to  experimentally determine the angular momentum of the drop has been to use Feynman's formula\cite{Fey55} to 
obtain the angular velocity $\omega$ from the vortex density, which  could be determined from the 2D droplet images, and the
above expression for $L_{\rm cap}$, see Ref. \onlinecite{Oco20} for details. 

\section{Summary}
\label{sec4}

We have shown that the presence of vortex arrays in  spinning  superfluid $^4$He droplets and
their interplay with capillary waves has a profound influence in the determination of  their globally stable configuration.
The presence of vortex arrays, irrespective of whether they are detected or not,\cite{Anc18} is the only plausible explanation for the existence of
oblate configurations,  which otherwise will be in conflict with quantum mechanics or would imply that droplets are not superfluid, which is extremely unlikely
due to the working temperatures in $^4$He droplets experiments.\cite{Toe04} 

Surface capillary waves are ubiquitous in prolate  superfluid $^4$He droplets.
Depending on their size and angular momentum, these waves may coexist with vortex arrays. We have found that 
the global equilibrium configuration of small prolate droplets is vortex-free, but the situation changes as the droplet size increases.		
This result is in agreement with a recent experiment,\cite{Oco20} where it has been disclosed that  vortex arrays and capillary waves coexist in 
very large drops.

While vortex arrays may be  detected by doping the droplets, capillary waves are
very elusive and escape direct detection. 
This poses a serious difficulty to  determine  the angular momentum of the superfluid droplet. 
From the theoretical side, microscopic approaches such as  DFT treat  vortex arrays and capillary waves  on the same footing,
and one has to resort to approximate methods to disentangle their contribution to the total angular momentum of the droplet. 
Determining --even in a model dependent way-- the angular momentum of the droplets is essential 
if one wants to characterize spinning superfluid droplets, as only  simultaneous knowledge of the droplet
morphology and angular momentum allows for a sensible comparison with classical or quantum models.

Contrarily to viscous droplets executing rigid body rotation (modeled by viscous droplets subject to surface tension and 
centrifugal forces {\it alone}),  the stability phase diagram  of superfluid droplets is not universal and cannot be characterized
by dimensionless angular momentum $\Lambda$ and dimensionless angular velocity $\Omega$ as its classical counterpart. Their knowledge  does not
determine  univocally the rotational properties of superfluid helium droplets, which display a clear dependence on the droplet size and/or
the number of vortices they host.
This is not in contradiction with the recent finding\cite{Oco20} that big superfluid $^4$He droplets hosting a large 
number of vortices seem to rotate like rigid-bodies, thus following the classical stability phase diagram. 
Rather, it is a manifestation  of finite size effects which still have to be studied in detail and, together with recent works on rotating $^3$He 
droplets\cite{Pi20,Ver20} and mixed $^3$He-$^4$He droplets,\cite{But20,Pib20} call for further experimental research. 

\begin{acknowledgments}
We are indebted to  Andrey Vilesov for useful discussions. We  thank  
Thomas M\"oller and Bruno Langbehn for providing us with the results of Ref. \onlinecite{Lan18}
shown in Fig.~\ref{fig1}, and  Sam Butler for providing us with the results of the classical model calculations used in this work.
This work has been 
performed under Grant No  FIS2017-87801-P (AEI/FEDER, UE).
J.M.E. acknowledges support from the Spanish Research Agency (AEI)
through the Severo Ochoa Centres of Excellence programme (grant SEV-2017-0706)
and from the European Union MaX Center of Excellence (EU-H2020 Grant No.\ 824143).
\end{acknowledgments}


\begin{thebibliography}{99}
\bibitem{note}
We remark that the term ``oblate'' will be used here to indicate droplets whose axial symmetry  is not appreciably distorted by 
the vortex array they host, as those appearing in the experiments. 
It is obvious that the presence of any vortex array breaks  axial symmetry, and {\it stricto sensu} these droplets are not  oblate.
\end{thebibliography}
\end{document}